\documentclass{iopart}
\usepackage{iopams,graphicx,graphics}

\begin{document}
\title{The gravitational-wave memory effect}
\author{Marc Favata\footnote{Present address: Theoretical Astrophysics, MC 350-17, California Institute of Technology, Pasadena, CA 91125}}
\address{Kavli Institute for Theoretical Physics, University of California, Santa Barbara, CA 93106}
\ead{favata@tapir.caltech.edu}
\begin{abstract}
The nonlinear memory effect is a slowly-growing, non-oscillatory contribution to the gravitational-wave amplitude. It originates from gravitational waves that are sourced by the previously emitted waves. In an ideal gravitational-wave interferometer a gravitational-wave with memory causes a permanent displacement of the test masses that persists after the wave has passed. Surprisingly, the nonlinear memory affects the signal amplitude starting at leading (Newtonian-quadrupole) order. Despite this fact, the nonlinear memory is not easily extracted from current numerical relativity simulations. After reviewing the linear and nonlinear memory I summarize some recent work, including: (1) computations of the memory contribution to the inspiral waveform amplitude (thus completing the waveform to third post-Newtonian order); (2) the first calculations of the nonlinear memory that include all phases of binary black hole coalescence (inspiral, merger, ringdown); and (3) realistic estimates of the detectability of the memory with LISA.
\end{abstract}
\section{Introduction}
We often think of gravitational-wave (GW) signals as having an oscillatory amplitude that starts small at early times, builds to some maximum, and then decays back to zero at late times. For example, this is the standard picture of a waveform from a coalescing compact-object binary. However, this picture is incomplete. In reality, \emph{all gravitational-wave sources} possess some form of \emph{gravitational-wave memory}. The GW signal from a `source with memory' has the property that the late-time and early-time values of at least one of the GW polarizations differ from zero:
\begin{equation}
\label{eq:memdef}
\Delta h_{+,\times}^{\rm mem} = \lim_{t\rightarrow +\infty} h_{+,\times}(t) - \lim_{t\rightarrow -\infty} h_{+,\times}(t),
\end{equation}
where $t$ is time at the observer.

When a GW \emph{without memory} passes through a detector, it causes oscillatory deformations but eventually returns the detector to its initial state. After a GW \emph{with memory} has passed through an \emph{idealized} detector (one that is truly freely-falling), it causes a permanent deformation---leaving a `memory' of the waves' passage. High-frequency detectors like bars or LIGO are rather insensitive to the memory from most sources because the detector response timescale is generally much shorter than the \emph{rise-time} of typical memory signals (the characteristic time for the non-oscillatory piece of the GW signal to build up to its final value). A detector like LISA is better able to detect the memory because of its good sensitivity in the low-frequency band where typical memory sources are stronger.\footnote{It has recently been realized that pulsar timing arrays \cite{sazhin-1978-pulsartiming,detweiler-ApJ1979-pulsartiming,hobbs-etal-amaldi8review} (which operate at much lower frequencies than LISA) could also be sensitive to the memory effect \cite{seto-memory-MNRAS09,levin-vanHaasteren-memory,pshirkov-etal-memory}, although the expected event rates from supermassive black hole binaries are small.}  Note also that bars and ground-based interferometers like LIGO are incapable of `storing' a memory signal because they are not truly free: internal elastic forces push a bar back to its equilibrium shape, and magnets on the LIGO test masses (as well as its pendulum wires) push them back to their equilibrium positions. Because its proof-masses are truly freely-floating, a detector like LISA could maintain a permanent displacement. However, this late-time displacement caused by the GW memory is not directly observable without information on the prior state of the detector: the spacetime metric near a detector long after a GW with memory has passed is equivalent to flat spacetime in non-standard coordinates (see Sec.~VD of \cite{favata-pnmemory} for details). Rather, it is the build-up of the memory (the difference in the metric between late and early times) that \emph{is} observable.

The memory effect has been known since the 1970's \cite{braginskii-grishchuk,zeldovich-polnarev,braginskii-thorne} in its linear form. The linear memory generally arises in systems with unbound components: a binary on a hyperbolic orbit (two-body scattering) \cite{turner-unbound}, matter or neutrinos ejected from a supernova \cite{epstein-neutrinomemory,turner-neutrinomemory,burrows-hayesPRL96}, or gamma-ray burst jets \cite{sago-GRBmemory}. In the 1990's a nonlinear form of memory was discovered independently by Blanchet \& Damour \cite{blanchet-damour-hereditary} and Christodoulou \cite{christodoulou-mem}. The nonlinear memory arises from the contribution of the emitted GWs to the changing quadrupole and higher mass moments. As discussed by Thorne \cite{kipmemory}, the nonlinear memory can be described in terms of a linear memory in which the unbound masses are the individual radiated gravitons. This implies that nearly all GW sources are sources with memory (even if the component objects remain bound).

Because the memory is a non-oscillatory effect with poor observational prospects for LIGO and other ground-based interferometers, it has been largely ignored by the GW community. But for the most important (or most studied) GW source---quasi-circular compact binaries---the nonlinear memory has quite a large contribution to the time-domain waveform amplitude: in a post-Newtonian (PN) expansion of the waveform polarizations, the memory effect enters at leading-(Newtonian)-order! That the memory enters at such low PN order is related to the fact that it is a \emph{hereditary} effect---the memory amplitude at any retarded time depends on the entire past motion of the source (and not just on the source's instantaneous retarded-time configuration). In addition, the nonlinear memory is a unique nonlinear effect because its non-oscillatory nature makes it distinctly visible in the waveform. For these reasons the memory should be studied further and its prospects for detection reassessed.

In the rest of this article I will briefly review the linear and nonlinear memory effects. I will then summarize my previous work in three areas: (i) computing the post-Newtonian memory corrections to the GW polarizations; (ii) calculating the evolution and saturation of the memory during the merger and ringdown of coalescing black hole (BH) binaries; and (iii) estimating the ability of LISA to detect the memory from supermassive BH binary mergers. Except for the presentation of the linear memory from hyperbolic binaries in Sec.~\ref{sec:linmem}, this conference proceeding concisely summarizes the results of references \cite{favata-pnmemory,favata-memory-saturation} as presented at the $8^{\rm th}$ Edoardo Amaldi Conference on Gravitational Waves. Readers are referred to those references for a more detailed exposition.

\section{\label{sec:prelim}Preliminaries}
We first introduce some formalism that will be useful for our discussion of the memory. The GW polarizations are conveniently decomposed in a sum over $(l,m)$ modes as
\begin{equation}
\label{eq:hdecompose}
h_{+} - \rmi h_{\times} = \sum_{l=2}^{\infty} \sum_{m=-l}^{l} h^{lm} {}_{-2}Y^{lm}(\Theta,\Phi) ,
\end{equation}
where ${}_{-2}Y^{lm}$ are spin-weighted spherical harmonics, and $(\Theta,\Phi)$ indicate the direction from the source to the observer. In a multipolar expansion of the GW field, the modes $h^{lm}$ are related to the radiative mass ($U^{lm}$) and current ($V^{lm}$) multipoles via
\begin{equation}
\label{eq:hlm}
h^{lm} = \frac{1}{\sqrt{2} R} \left[ U^{lm}(T_R) - \rmi V^{lm}(T_R) \right].
\end{equation}
Here $R$ is the distance from source to observer, $T_R$ is retarded time, our units assume $G=c=1$, and our overall sign and normalization depends on our polarization triad and other conventions (see \cite{favata-pnmemory} for details). The moments $U^{lm}$ and $V^{lm}$ are constructed from their corresponding symmetric-trace-free (STF) tensors of rank-$l$:
\begin{equation}
\label{eq:Ulmdef}
U^{lm} = A_l \, {\mathcal U}_L {\mathcal Y}^{lm \, \ast}_L \qquad {\rm{and}} \qquad V^{lm} = B_l \, {\mathcal V}_L {\mathcal Y}^{lm \, \ast}_L,
\end{equation}
where $A_l$ and $B_l$ are $l$-dependent constants and ${\mathcal Y}^{lm}_L$ are the STF spherical harmonics.\footnote{These are related to the ordinary scalar spherical harmonics via $Y^{lm} = {\mathcal Y}^{lm}_L n_L$, where $n_L=n_{i_1} n_{i_2} \cdots n_{i_l}$ is a product of $l$ unit radial vectors. On the quantities ${\mathcal U}_L$, ${\mathcal V}_L$, and ${\mathcal Y}^{lm}_L$, $L$ represents $l$ tensor indices (e.g., for $l=3$, ${\mathcal U}_L \rightarrow {\mathcal U}_{abc}$).}

The radiative moments $U^{lm}$ and $V^{lm}$ that appear in the wavezone expansion of $h_{+,\times}$ are related to a family of \emph{source multipole moments} ($I_{lm},\, J_{lm},\, \ldots$). These source moments are defined in terms of integrals over the stress-energy pseudotensor of the matter and gravitational fields of the source through a \emph{multipolar post-Minkowski} iteration scheme (see \cite{blanchetLRR} for a review or Sec.~II of \cite{favata-pnmemory} for a brief summary). For example, the radiative mass moments are related to the source mass moments via
\begin{equation}
\label{eq:Ulmrelate}
\fl U_{lm} = I_{lm}^{(l)} + 2 {\mathcal M} \int_{-\infty}^{T_R} \left[ \ln\left(\frac{T_R - \tau}{2\tau_0}\right) + \kappa_l  \right] I_{lm}^{(l+2)}(\tau) \rmd\tau + U_{lm}^{({\rm{nonlin mem}})} + \Or(2.5{\rm PN}).
\end{equation}
Here $I_{lm}^{(l)}$ is the $l^{\rm th}$ time derivative of the mass source moment $I_{lm}$, the integral term is a 1.5PN order tail term, ${\mathcal M}$ is the mass monopole moment, $\kappa_l$ is an $l$-dependent constant,  $\tau_0$ is an arbitrary timescale that disappears in physical observables,  $U_{lm}^{({\rm{nonlin mem}})}$ is the nonlinear memory term (discussed below), and $\Or(2.5{\rm PN})$ refers to several types of terms that enter at 2.5PN and higher orders.
\section{\label{sec:linmem}Linear memory}
\begin{figure*}[t]
$
\begin{array}{cc}
\includegraphics[angle=0, width=0.46\textwidth]{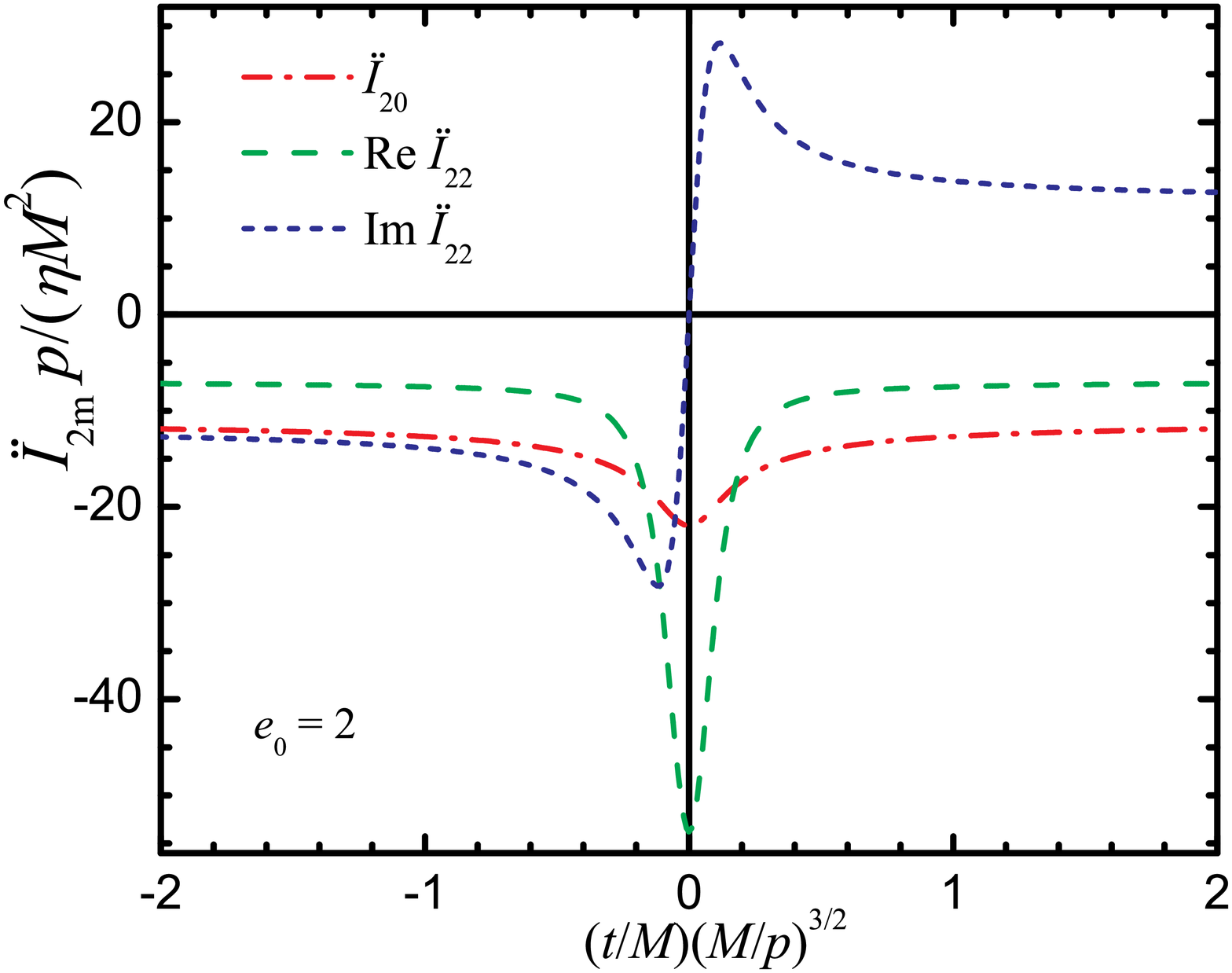} &
\includegraphics[angle=0, width=0.48\textwidth]{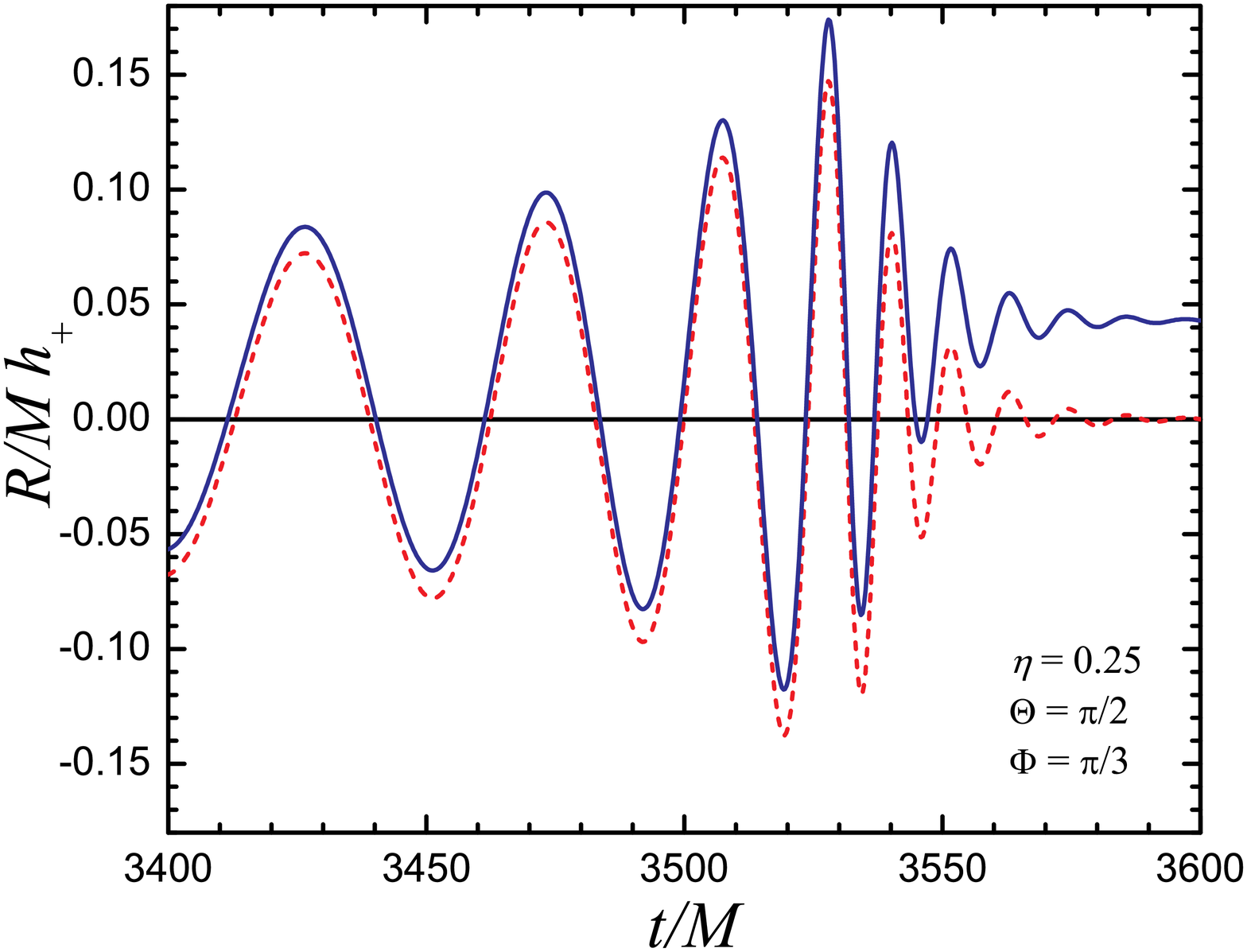}
\end{array}
$
\caption{\label{fig:hmem}Examples of gravitational-wave signals with memory. The left plot shows the waveform modes $\ddot{I}_{2m}$ for a hyperbolic orbit with eccentricity $e_0=2$ as a function of time [see (\ref{eq:Ilmhyperbolic}) and surrounding text]. Note the linear memory present in the imaginary part of $\ddot{I}_{22}$. The right plot shows the $h_{+}$ polarization for an equal-mass binary black hole coalescence with (blue/solid) and without (red/dashed) the nonlinear memory. The oscillatory piece of $h_+$ was computed using an effective-one-body (EOB) model for the $h_{22}$ mode. The memory piece was computed by substituting this mode into (\ref{eq:memhplus}). See \cite{favata-memory-saturation} for details.}
\end{figure*}

As a simple example of the linear memory, let's consider the waveform from a hyperbolic binary. The leading-order multipolar contribution to the polarizations is
\begin{equation}
\label{eq:hkepler}
h_+ - \rmi h_{\times} \approx \sum_{m=-2}^{2} \frac{I_{2m}^{(2)}}{R \sqrt{2}} {}_{-2}Y^{lm}(\Theta,\Phi).
\end{equation}
For a Keplerian binary in the $x$-$y$ plane with relative orbital separation $r(t)$, total mass $M=m_1+m_2$, reduced mass ratio $\eta\equiv m_1 m_2/M^2$, and orbital phase angle $\varphi(t)$, the mass quadrupole is $I_{2m}=(16\pi/5\sqrt{3}) \eta M r(t)^2 Y_{2m}^{\ast}(\pi/2,\varphi(t))$. For Keplerian orbits with semi-latus rectum $p$, eccentricity $e_0$, and true anomaly $v=\varphi - \omega_p$ ($\omega_p=0$ sets the periastron direction on the $x$-axis), the orbital motion is described by
\begin{equation}
\label{eq:keplerianeqns}
r=\frac{p}{1+e_0 \cos v} \qquad {\rm and} \qquad  \dot{v} = \dot{\varphi} = \frac{\sqrt{pM}}{r^2}.
\end{equation}
The resulting waveforms are given by (\ref{eq:hkepler}) and
\begin{eqnarray}
\label{eq:Ilmhyperbolic}
I_{20}^{(2)} &= -8 \sqrt{\frac{\pi}{15}} \eta \frac{M^2}{p} e_0 (e_0 + \cos v), \\
I_{2 \pm 2}^{(2)} &=-4 \sqrt{\frac{2\pi}{5}} \eta \frac{M^2}{p} \rme^{\mp 2 \rmi \varphi(t)} \left[ 1-e_0^2 + (1+e_0 \cos v)(1+2 e_0 \rme^{\pm \rmi v}) \right].
\end{eqnarray}
For $0 \leq e_0 < 1$ these waveforms are clearly oscillatory. But for a hyperbolic orbit ($e_0>1$, with $\omega_p=0$) the phase angle approaches $\varphi_{-}=v_{-}=-\arccos(-e_0^{-1})$ at early times ($t\rightarrow -\infty$), while at late times ($t\rightarrow +\infty$) it approaches $\varphi_{+}=v_{+}=\arccos(-e_0^{-1})$. This difference in the late and early time values of the orbital phase angle yields a corresponding difference in the derivatives of the mass multipoles,
\numparts
\begin{eqnarray}
\Delta I_{20}^{(2)} &=0,\\
\Delta I_{2\pm2}^{(2)} &= \pm \rmi 16 \sqrt{\frac{2\pi}{5}} \frac{\eta M^2}{p} \frac{(e_0^2-1)^{3/2}}{e_0^2},
\end{eqnarray}
\endnumparts
resulting in a memory in the GW polarization amplitudes (see figure \ref{fig:hmem}). Note that for a parabolic orbit ($e_0=1$) there is no memory since the orbital phase angle returns to its early-time value. As in the above example, the linear memory always arises from a change in the mass or current \emph{source} moment derivatives, $\Delta I_{lm}^{(l)}$ or $\Delta J_{lm}^{(l)}$ for $l \geq 2$.

We can also derive the linear memory for an unbound system by solving the linearized, harmonic gauge Einstein field equations (EFE) for the space-space piece of the metric perturbation $h_{jk}$: $\Box \bar{h}_{jk} = -16 \pi T_{jk}$. Here $T_{jk}$ is the stress-energy tensor of $N$ gravitationally unbound particles with masses $M_A$ and constant velocities ${\bi{v}}_A$, $\bar{h}_{jk}$ is the trace-reversed metric perturbation, and $\Box$ is the flat-space wave operator. Solving this equation (via the Li\'{e}nard-Wiechert solution) and projecting to transverse-traceless (TT) gauge yields \cite{braginskii-thorne,kipmemory}:
\begin{equation}
\label{eq:hijlinmem}
\Delta h_{jk}^{\rm TT} = \Delta \sum_{A=1}^N \frac{4 M_A}{R\sqrt{1-v_A^2}} \left[ \frac{v_A^j v_A^k}{1-{\bi{v}}_A \cdot {\bi{N}}} \right]^{\rm TT}.
\end{equation}
Here ${\bi{N}}$ points from the source to the observer and $\Delta$ means to take the difference between the late and early time values of the summation. In this formula the masses and velocities could refer to (i) the masses and velocities of the pieces of a disrupted binary, (ii) a gamma-ray-burst jet \cite{sago-GRBmemory}, or (iii) the individual radiated neutrinos \cite{epstein-neutrinomemory,turner-neutrinomemory} or pieces of ejected material in a supernova explosion (see, e.g., \cite{burrows-hayesPRL96,ott-corecollapsereview,ott-murphy-burrows-2009} for numerical simulations of supernovae that show a memory effect.)
\section{\label{sec:nonlinmem}Nonlinear (Christodoulou) memory}
The \emph{nonlinear memory} \cite{payne-zfl,blanchet-damour-hereditary,christodoulou-mem} (often referred to as the `Christodoulou memory') arises from a contribution to the radiative mass multipole moments $U_{lm}$ that is sourced by the energy-flux of the radiated GWs. Consider the relaxed EFE in harmonic gauge: $\Box \bar{h}^{\alpha \beta} = -16 \pi \tau^{\alpha \beta}$, where $\tau^{\alpha \beta}$ depends on the matter stress-energy tensor $T^{\alpha \beta}$, the Landau-Lifshitz pseudotensor $t_{\rm LL}^{\alpha \beta}$, and other terms quadratic in $\bar{h}^{\alpha \beta}$ \cite{kiprmp}. Of the many nonlinear terms in $t_{\rm LL}^{\alpha \beta}$, there is a piece that is proportional to the stress-energy tensor for GWs:
$T^{\rm gw}_{jk} = \frac{1}{R^2} \frac{\rmd E^{\rm gw}}{\rmd t \rmd \Omega} n_j n_k$,
where $\frac{dE^{\rm gw}}{dt d\Omega}$ is the GW energy flux and $n_j$ is a unit radial vector.  When applying the Green's function $\Box^{-1}$ to the right-hand-side of the EFE, this piece yields the following correction term to the GW field \cite{wiseman-will-memory}:
\begin{equation}
\label{eq:hTTnonlinmem}
\delta h^{\rm TT}_{jk} = \frac{4}{R} \int_{-\infty}^{T_R} \rmd t'\, \left[ \int \frac{\rmd E^{\rm gw}}{\rmd t' \rmd \Omega'} \frac{n'_j n'_k}{(1-{\bi{n}}' \cdot {\bi{N}})} \rmd \Omega' \right]^{\rm TT},
\end{equation}
where $T_R$ is the retarded time. The time-integral in the above equation is what gives the memory its hereditary nature: the memory piece of the GW field for any value of $T_R$ depends on the entire past history of the source.
Thorne \cite{kipmemory} has shown that the nonlinear memory (\ref{eq:hTTnonlinmem}) can be described in terms of the linear memory (\ref{eq:hijlinmem}) if the unbound objects in the system are taken to be the individual radiated gravitons with energies $E_A=M_A/(1-v_A^2)^{1/2}$ and velocities $v^j_A = c \, n_{A}'^j$.

When the GW field is decomposed into modes as in (\ref{eq:hdecompose}), the nonlinear memory can be shown to yield a correction to the radiative mass multipole moments (\ref{eq:Ulmrelate}) that enters at 2.5PN and higher orders \cite{blanchet-damour-hereditary,favata-pnmemory}:
\begin{equation}
\label{eq:Ulmmem}
U_{lm}^{\rm (nonlin mem)} = 32\pi \sqrt{\frac{(l-2)!}{2(l+2)!}} \int_{-\infty}^{T_R} \!\! \rmd t \int \! \rmd \Omega \, \frac{\rmd E_{\rm gw}}{\rmd t \rmd \Omega}(\Omega) Y_{lm}^{\ast}(\Omega) .
\end{equation}
The radiative current moments $V_{lm}$ do not have a nonlinear memory contribution.
Note that the energy flux is itself defined in terms of the waveform modes via
\begin{equation}
\label{eq:dEdtdOmega-hlm}
\fl \frac{\rmd E_{\rm gw}}{\rmd t \rmd \Omega} =\frac{R^2}{16\pi} \langle \dot{h}_{+}^2 + \dot{h}_{\times}^2 \rangle
= \frac{R^2}{16\pi} \sum_{l', l'', m', m''} \langle \dot{h}_{l'm'} \dot{h}^{\ast}_{l''m''} \rangle {}_{-2}Y^{l' m'}(\Omega) {}_{-2}Y^{l'' m'' \,\ast}(\Omega).
\end{equation}
This means that the memory is calculated in an iterative fashion: those $h_{lm}$ modes that contain the memory are negligible in the computation of the energy flux.

The memory (linear and nonlinear) should not be mistaken as a change in the monopolar-piece of the $1/R$ expansion of the metric.  Rather it is a change in the quadrupolar (and higher-order) pieces of the $1/R$-spatial-part of the TT-projection of the metric. It is a purely GW effect, and is not directly connected with the change in the `Coulomb part' of the metric or the `mass loss of the source' (spherically symmetric mass loss produces no GWs!) except indirectly through the changing mass's effect on the quadrupole and higher-order multipole moments.

\section{\label{sec:PNmemory}Memory contribution to the post-Newtonian waveform of quasi-circular, inspiralling binaries}
To compute the nonlinear memory contribution to the waveform for quasi-circular binaries, one begins by substituting (\ref{eq:dEdtdOmega-hlm}) into (\ref{eq:Ulmmem}) and evaluating the angular integral in (\ref{eq:Ulmmem}). Next one substitutes explicit expressions for the $h_{lm}$ modes into the result. These modes are proportional to $h_{lm} \propto \rme^{-\rmi m\varphi(t)}$ and are given explicitly (to 3PN order and for $m\neq 0$) in \cite{blanchet3pnwaveform}. The resulting integrand of the time-integral in (\ref{eq:Ulmmem}) will be either oscillatory or non-oscillatory. The non-vanishing, non-oscillatory modes turn out to be the $U_{l0}^{({\rm{nonlin mem}})}$ with $l$-even. (The $m\neq 0$ modes are oscillatory---they do not contribute to the memory, but they affect the waveform at 2.5PN and higher orders.) These $U_{l0}^{({\rm{nonlin mem}})}$ consist of time-integrals of polynomials in $x\equiv (M\omega)^{2/3}$ (where $\omega$ is the orbital angular frequency), with each term of the form
\begin{equation}
\label{eq:memoryint}
\int_{-\infty}^{T_R} [x(t)]^n \, \rmd t = \int_{0}^{x(T_R)} \frac{x^n}{\dot{x}} \, \rmd x .
\end{equation}
After performing the change of variable indicated in (\ref{eq:memoryint}), evaluation of the resulting integrals requires a model for the frequency evolution of the binary. The adiabatic evolution of the frequency (or $x$) is easily derived from energy balance (GW luminosity equals the rate of change of the orbital energy, ${\mathcal L}_{\rm GW} = -\dot{E}$) and the relation $\dot{x} = -{\mathcal L}_{\rm GW}/(\rmd E/\rmd x)$. The resulting frequency evolution (known to 3.5PN order),
\begin{equation}
\frac{\rmd x}{\rmd t} = \frac{64}{5} \frac{\eta}{M} x^5 \left[ 1 +\Or(x) + \Or(x^{3/2}) + \cdots + \Or(x^{7/2}) + \Or(4{\rm PN}) \right],
\end{equation}
can be used to evaluate integrals of the form (\ref{eq:memoryint}). For details see \cite{favata-pnmemory}.

When the above procedure is carried out, the resulting memory contributions only affect the $+$ waveform polarizations\footnote{This is a consequence of our choice for the polarization triad. A rotation of the triad would cause memory in both polarizations.} and have the form
\begin{equation}
h_{+}^{\rm (mem)} = \frac{2\eta M x}{R} \sum_{n=0}^{\infty} x^{n/2} H_+^{(n/2, {\rm mem})}, \;\;{\rm where}
\end{equation}
\numparts
\label{eq:PNHplusmem}
\begin{equation}
\label{eq:Hplus0pn}
\fl H_{+}^{(0, {\rm mem})} =\frac{1}{96} s^2_{\Theta} (17 + c^2_{\Theta} ) ,
\end{equation}
\begin{equation}
\label{eq:Hplus05pn}
\fl H_{+}^{(0.5, {\rm mem})} = 0 = H_{+}^{(1.5, {\rm mem})},
\end{equation}
\begin{eqnarray}
\label{eq:Hplus1pn}
\fl H_{+}^{(1, {\rm mem})} = s^2_{\Theta} \left[ - \frac{354\,241}{2\,064\,384} - \frac{62\,059}{1\,032\,192} c^2_{\Theta} - \frac{4195}{688\,128} c^4_{\Theta}
\right. \nonumber \\ \left.
+ \left( \frac{15\,607}{73\,728} + \frac{9373}{36\,864} c^2_{\Theta} + \frac{215}{8192} c^4_{\Theta} \right) \eta \right] ,
\end{eqnarray}
\begin{eqnarray}
\label{eq:Hplus2pn}
\fl H_{+}^{(2, {\rm mem})} =  s^2_{\Theta} \left[ - \frac{3\,968\,456\,539}{9\,364\,045\,824} + \frac{570\,408\,173}{4\,682\,022\,912} c^2_{\Theta} + \frac{122\,166\,887}{3\,121\,348\,608} c^4_{\Theta} + \frac{75\,601}{15\,925\,248} c^6_{\Theta}
\right. \nonumber \\
+ \left( - \frac{7\,169\,749}{18\,579\,456} - \frac{13\,220\,477}{18\,579\,456} c^2_{\Theta} - \frac{1\,345\,405}{6\,193\,152} c^4_{\Theta}  - \frac{25\,115}{884\,736} c^6_{\Theta} \right) \eta
\nonumber \\ \left.
+ \left(  \frac{10\,097}{147\,456} + \frac{5179}{36\,864} c^2_{\Theta} + \frac{44\,765}{147\,456} c^4_{\Theta} + \frac{3395}{73\,728} c^6_{\Theta} \right) \eta^2 \right] ,
\end{eqnarray}
\begin{equation}
\label{eq:Hplus25pn}
\fl H_{+}^{(2.5, {\rm mem})} = - \frac{5\pi}{21\,504} (1-4\eta) s^2_{\Theta} \left( 509 + 472 c^2_{\Theta} + 39 c^4_{\Theta} \right),
\end{equation}
\begin{equation}
\label{eq:Hplus3pn}
\fl H_{+}^{(3, {\rm mem})} = {\rm Equation\; (4.6g)\; of\;} \cite{favata-pnmemory},
\end{equation}
\endnumparts
where $c_{\Theta} \equiv \cos\Theta$ and $s_{\Theta}\equiv \sin\Theta$.

The 0PN memory contribution (\ref{eq:Hplus0pn}) was first computed in \cite{wiseman-will-memory}; the 0.5PN term was computed in \cite{blanchet3pnwaveform}. All the other PN corrections listed above are new and derived in \cite{favata-pnmemory}. Combined with the oscillatory waveform pieces computed in \cite{blanchet3pnwaveform}, these results complete the waveform polarizations at 3PN order.

Note in particular that at Newtonian (0PN) order, the total waveform is given by
\numparts
\begin{equation}
\label{eq:hplusquad}
\fl h_{+}^{({0\rm PN})} =  2 \frac{\eta M}{R} x \left[ -(1+c^2_{\Theta}) \cos2(\varphi + \Phi) +\frac{1}{96} s^2_{\Theta} (17+c^2_{\Theta}) + \Or(x^{1/2}) \right],
\end{equation}
\begin{equation}
\fl h_{\times}^{(0\rm PN)} =  2 \frac{\eta M}{R} x \left[ -2 c_{\Theta} \sin2(\varphi - \Phi) + \Or(x^{1/2}) \right].
\end{equation}
\endnumparts
This shows that even though the memory originates in a high-PN-order effect, it enters the waveform at the same order as the standard, leading-order `quadrupole' piece. This fact is not widely appreciated.
\section{\label{sec:mergermem}Computing the memory in binary black hole mergers}
\subsection{\label{sec:nrlimits}Limitations of numerical relativity in modeling the memory}
The PN waveforms computed above allow us to model the slow growth of the memory during the inspiral phase of coalescence. To determine the rest of the memory's evolution and its eventual saturation value, we need to model the merger and ringdown phases. Numerical relativity (NR) provides the only accurate way to do this. Unfortunately, the memory is not a quantity that is easily computed by current NR simulations. Indeed, one may have noticed that none of the waveforms produced by recent NR simulations show any sign of memory.

There are several reasons for this. First, for quasi-circular inspiralling binaries the memory is not present in the commonly plotted $l=m=2$ mode. The memory is only present in the $m=0$ modes of the waveform. More importantly, while the memory in the metric perturbation and its $h_{l0}$ modes can be comparable in size to the leading-order oscillatory $h_{2m}$ modes [see e.g., (\ref{eq:hplusquad})], the memory contribution to $\Psi_4$ (the Weyl scalar that NR simulations usually use to extract GWs) is suppressed by several orders-of-magnitude relative to the leading-order oscillatory piece of $\Psi_4$. For example at large $R$, $\Psi_4 \rightarrow \ddot{h}_+ - \rmi \ddot{h}_{\times}$, and a decomposition of $\Psi_4$ similar to (\ref{eq:hdecompose}) implies that its modes $\psi_{lm}$ are related to the metric perturbation modes via $\psi_{lm}=\ddot{h}_{lm}$. During the inspiral the metric modes be can decomposed into an amplitude $A_{lm}(t)$ that slowly evolves on a radiation-reaction timescale $T_{\rm rr}\propto (M/\eta)(M\omega)^{-8/3}$, and a phase that oscillates on an orbital timescale $T_{\rm orb}\propto 1/\omega$: $h_{lm}=A_{lm}(t) \rme^{-\rmi m\omega t}$. Then for oscillatory ($m\neq 0$) modes  $\psi_{lm}$ is smaller by two factors of the orbital time, $\psi_{lm} \sim \omega^2 h_{lm}$, but for memory ($m=0$) modes  $\psi_{l0}$ is smaller by two factors of the radiation-reaction time, $\psi_{l0} \sim \eta^2/M^2 (M\omega)^{16/3} h_{l0}$. This means that the $\psi_{l0}$ memory modes are smaller than the oscillatory $\psi_{lm}$ modes by 5PN orders. The memory modes of $\Psi_4$ are therefore a small correction superimposed on the much larger oscillatory pieces of $\Psi_4$ and are thus difficult to resolve in a numerical simulation.

Aside from this problem, there is also the issue of computing the two integration constants needed to go from $\psi_{lm}$ to $h_{lm}$. Berti et.~al \cite{berti-etal-multipolarnonspinning} showed that choosing these constants incorrectly leads to an `artificial memory' that shows up in all ($l,m$) modes. One approach to computing these constants correctly is to match the NR waveform modes onto the PN waveform modes.

Some of the above problems can be ameliorated by using a wave-extraction formalism that directly computes the metric waveform without the need to compute $\Psi_4$ and integrate twice. However another problem arises that also affects metric-extraction schemes: the sensitivity of the nonlinear memory to the distant past requires large initial binary separations ($\gtrsim 50M$) to compute the $h_{l0}$ modes accurately. See Sec.~VC of \cite{favata-pnmemory} for further discussion of these issues.
\subsection{\label{sec:mmwmodel}Analytic modeling of the full coalescence}
Since NR simulations of the coalescence memory are not yet available, we need to investigate other approaches. A first attempt by Kennefick \cite{kennefick-memory} took the leading-order inspiral memory [(\ref{eq:hplusquad}), which grows as $h_+^{\rm mem, insp}\propto (t_c-t)^{-1/4}$, where $t_c$ is the coalescence time] and assumed that it abruptly stops growing at a time corresponding to a specified (and somewhat arbitrary) orbital separation (roughly the last-stable-orbit). This approach truncates the memory too early and does not incorporate its smooth build-up during the merger and ringdown.

To improve upon Kennefick's analysis we begin by introducing the so-called \emph{minimal waveform model} (MWM). The purpose of the MWM is to use information from NR and the effective-one-body (EOB) \cite{EOB-damour-lecnotes} approach to construct a very simple, fully-analytic waveform that qualitatively contains the basic features of a binary BH coalescence waveform. The MWM simply consists of matching the leading-order inspiral waveform to a ringdown waveform. For example, the $q^{\rm th}$ time-derivative of the $(2,2)$ mode is approximated as
\begin{equation}
\label{eq:MWM22}
\fl h_{2\pm2}^{(q)} \approx \frac{I_{2\pm2}^{(q+2)}}{R\sqrt{2}} = \frac{1}{R\sqrt{2}} \times \cases{2\sqrt{\frac{2\pi}{5}}\eta M r^2 (\mp 2\rmi \omega)^{q+2} \rme^{\mp 2\rmi \varphi}&for $t\leq t_m$\\
\sum_{n=0}^{n_{\rm max}} (-\sigma_{22n})^q A_{22n} \rme^{-\sigma_{22n} (t-t_m)}&for $t>t_m,$}
\end{equation}
where $\omega\equiv \dot{\varphi}=(M/r^3)^{1/2}$, $r=r_m [1-(t-t_m)/\tau_{\rm rr}]^{1/4}$, $\tau_{\rm rr}=(5/256)(M/\eta)(r_m/M)^4$, $\sigma_{lmn}=\rmi\omega_{lmn} + \tau^{-1}_{lmn}$ are the BH quasi-normal mode frequencies and damping times \cite{berti-cardoso-will-PRD2006} (these depend on the final BH mass and spin which are determined from NR simulations), and $t_m$ is a `matching time' at which $r=r_m$. The matching radius $r_m$ is an adjustable parameter that determines the peak of the waveform amplitude; I choose the value $r_m=3M$ corresponding to the Schwarzschild light-ring. The coefficients $A_{lmn}$ are determined by matching (\ref{eq:MWM22}) at $t=t_m$ for $0 \leq q \leq n_{\rm max}$.

For quasi-circular orbits, the leading-order piece of the memory is sourced by the $h_{22}$ oscillatory waveform mode. To compute its contribution to the memory, substitute the $l=2$ expansion of the energy flux (\ref{eq:dEdtdOmega-hlm}) into (\ref{eq:Ulmmem}) to yield $U_{l0}^{\rm (nonlin mem)}$ and the corresponding $h_{l0}$ modes. The resulting memory contribution to $h_+$ is
\begin{equation}
\label{eq:memhplus}
h_{+}^{\rm (mem)} \approx \frac{R}{192\pi} s_{\Theta}^2 (17+c_{\Theta}^2) \int_{-\infty}^{T_R} |\dot{h}_{22}|^2 \rmd t.
\end{equation}
This reduces to a simple analytic expression when the MWM (\ref{eq:MWM22}) is substituted (see \cite{favata-memory-saturation} for details). The effect of including higher-order waveform modes in (\ref{eq:dEdtdOmega-hlm}) is currently under investigation. Because the memory (\ref{eq:memhplus}) depends only on the amplitude of the $h_{22}$ mode and not its phase, the MWM provides a reasonably accurate description of the memory. As an alternative to this simple analytic model, one can use an EOB description of the $h_{22}$ mode and substitute this into (\ref{eq:memhplus}). This approach is more accurate because the EOB waveforms are calibrated to NR simulations. Both methods were implemented and compared in \cite{favata-memory-saturation}. Figure \ref{fig:hmem} shows the memory resulting from the EOB-based model.
\section{\label{sec:memdetect}Detecting the nonlinear memory}
Although the memory has most of its power at near-zero frequency, the build-up of the memory distributes power in a range of frequencies. As a simple estimate of the memory's detectability with GW interferometers, we compute the sky-averaged rms signal-to-noise ratio (SNR) for a detector with noise spectral density $S_n(f)$,
\begin{equation}
{\rm SNR} = \left[ \int_{0}^{\infty} \frac{h_c^2(f)}{h_n^2(f)} \frac{\rmd f}{f} \right]^{1/2},
\end{equation}
where the sky-averaged rms noise amplitude is $h_n(f)=\sqrt{\alpha f S_n(f)}$ ($\alpha=5$ for LIGO-like interferometers, $20/3$ for LISA), and the characteristic memory amplitude is
\begin{equation}
h_c(f)=2 (1+z) f \langle | \tilde{h}_{+}^{\rm (mem)}[(1+z)f] |^2 \rangle^{1/2}|_{R\rightarrow D_L/(1+z)},
\end{equation}
where $\tilde{h}_{+}^{\rm (mem)}$ is the Fourier transform of $h_{+}^{\rm (mem)}$, $D_L(z)$ is the luminosity distance, and the angle brackets mean to average over sky-position and polarization angles. For the MWM the Fourier transform for $f>0$ can be expressed analytically as \cite{favata-memory-saturation}
\begin{eqnarray}
\label{eq:FThmem}
\fl \tilde{h}_{+}^{\rm (mem, MWM)}(f) = \frac{\eta M}{384 \pi R} s^2_{\Theta} (17+c^2_{\Theta}) \frac{\rmi {\mathcal F}}{2\pi f} \Bigg\{ \frac{8\pi M}{r_m} \left[ 1- 2\pi \rmi f \tau_{\rm rr} U(1,7/4,2\pi \rmi f \tau_{\rm rr}) \right]
\nonumber \\
  - \frac{1}{\eta M}  \sum_{n,n'=0}^{n_{\rm max}} \frac{\sigma_{22n}^{\,} \sigma_{22n'}^{\ast} A_{22n}^{\,} A_{22n'}^{\ast}}{2\pi \rmi f - (\sigma_{22n}^{\,} + \sigma_{22n'}^{\ast})} \Bigg\}.
\end{eqnarray}
where $U$ is Kummer's confluent hypergeometric function of the second kind and ${\mathcal F}$ is a `fudge factor' that matches the MWM and EOB memory models at late times.

\begin{figure*}[t]
$
\begin{array}{cc}
\includegraphics[angle=0, width=0.48\textwidth]{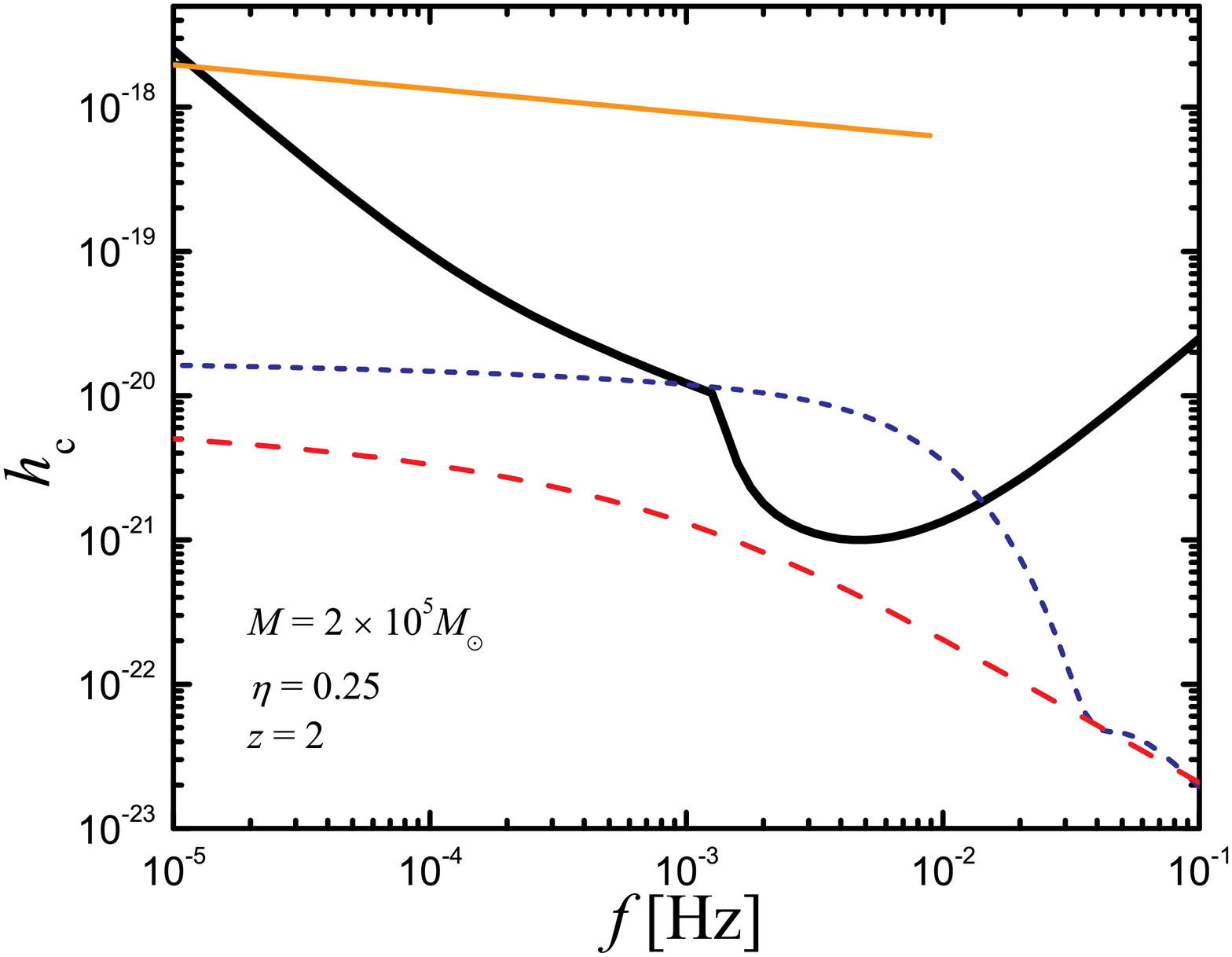} &
\includegraphics[angle=0, width=0.46\textwidth]{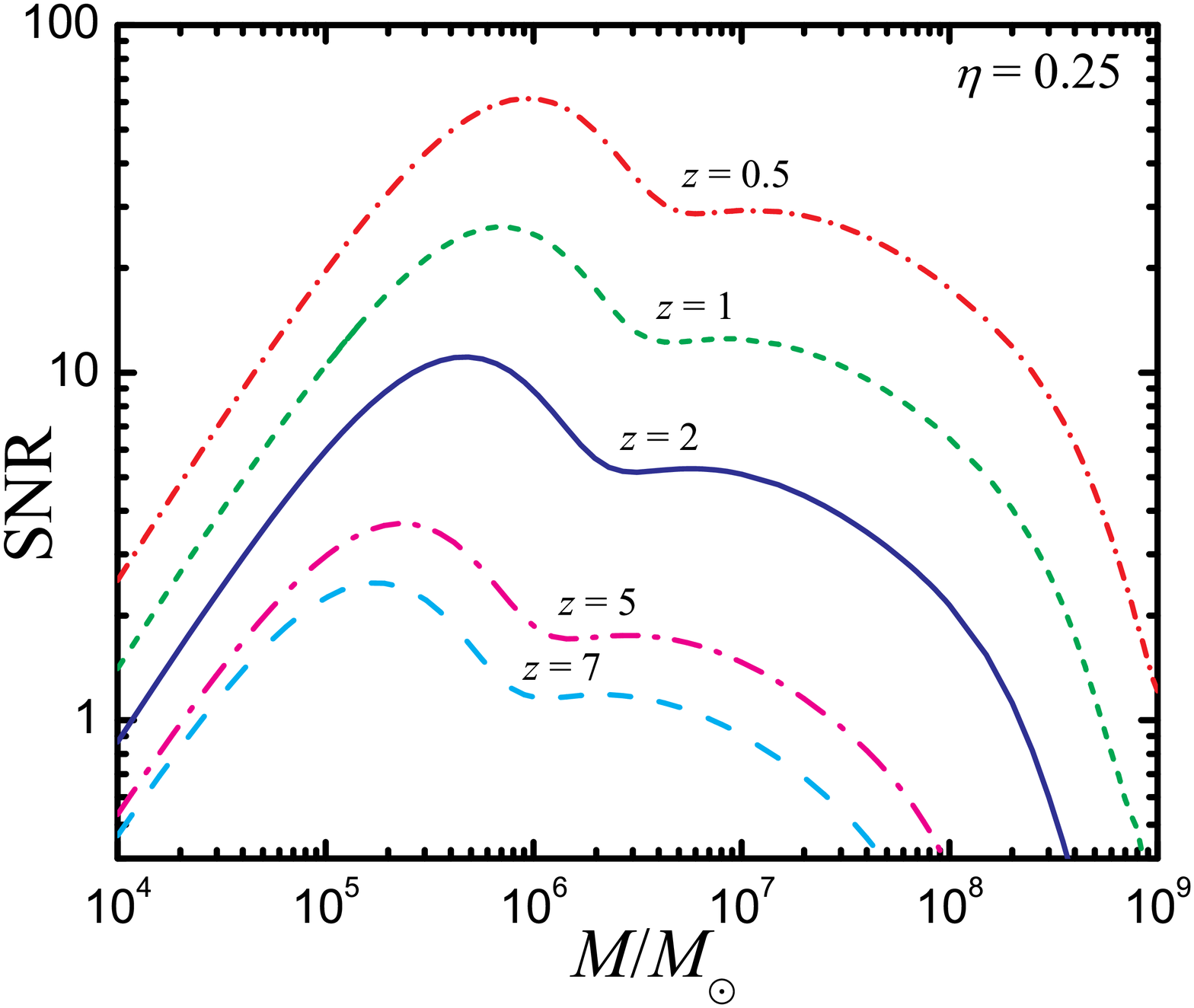}
\end{array}
$
\caption{\label{fig:memdetect}Detectability of the memory with LISA. The left plot shows the LISA noise amplitude $h_n$ (solid/black) and the characteristic amplitudes $h_c$ for the inspiral waves (solid/orange) and the nonlinear memory for a black hole binary with the indicated parameters. The short-dashed/blue curve uses the MWM (\ref{eq:FThmem}) to compute the memory; the long-dashed/red curve uses a truncated-inspiral model based on \cite{kennefick-memory} (see \cite{favata-memory-saturation} for details). The right plot shows the angle-averaged SNR of the nonlinear memory signal for equal-mass LISA binaries as a function of the total binary (source-frame) mass $M$ and redshift $z$.}
\end{figure*}

Evaluating the SNR, one finds that initial LIGO can only detect the memory from stellar-mass BH mergers within the Local Group. Advanced LIGO will see the memory to about 10 times that distance ($\lesssim 20$ Mpc). LISA, however, will see the memory from supermassive BH mergers out to redshifts $z\lesssim 2$; see Figure \ref{fig:memdetect}.
\section{\label{sec:concl}Conclusions}
Gravitational waves with memory are potential sources for detectors with good low-frequency sensitivity (LISA and pulsar-timing arrays) and are worth further study. The nonlinear memory is particularly interesting because (i) it arises from the gravitational waves produced by gravitational waves, (ii) it affects the waveform at leading-order, and (iii) it imparts a unique and visually apparent signature to the waveform. Although our knowledge of the oscillatory pieces of the waveform has greatly increased in recent years (thanks to the work of many in the PN and NR communities), our understanding of the memory has not. Here I summarized recent attempts to fill in the gaps in our understanding. This included determining the missing pieces of the 3PN quasi-circular inspiral waveform that arise from the memory. It also included a first study of the build up and saturation value of the memory during the merger and ringdown phases. As discussed above, numerical relativity simulations cannot yet easily compute the nonlinear memory. This study therefore serves as an example of how post-Newtonian and effective-one-body methods, combined with crucial input from numerical relativity simulations, allows us to compute quantities that are not directly computable with analytical or numerical techniques alone. The prospects for detecting the nonlinear memory were also investigated and found to be somewhat poor for advanced LIGO but good for LISA. Motivated by this work, several other studies \cite{seto-memory-MNRAS09,levin-vanHaasteren-memory,pshirkov-etal-memory} have recently investigated the detectability of the memory with pulsar timing arrays. Work is in progress to improve upon the initial estimates of the merger/ringdown memory discussed here, extend some results to eccentric orbits, and perform a more detailed study of the memory's detectability with LISA and future ground-based detectors.

\ack
This research was supported by the National Science Foundation under
Grant No. PHY05-51164 to the Kavli Institute for Theoretical Physics, and by the NASA Postdoctoral Program at the Jet Propulsion Laboratory.


\end{document}